\documentclass[12pt,epsf,amssymb]{article} \textheight =23 truecm
\def\lsim{\raise0.3ex\hbox{$<$\kern-0.75em\raise-1.1ex\hbox{$\sim$}}}
\def\gsim{\raise0.3ex\hbox{$>$\kern-0.75em\raise-1.1ex\hbox{$\sim$}}}
\def\noi{\noindent} \def\nn{\nonumber} \def\bea{\begin{eqnarray}}
\def\eea{\end{eqnarray}} \def\beq{\begin{equation}}
\def\eeq{\end{equation}} 
\def\beeq{\begin{eqnarray}} \def\eeeq{\end{eqnarray}} \def\R{ {\rm R
\kern -.31cm I \kern .15cm}} \def\C{ {\rm C \kern -.15cm \vrule
width.5pt \kern .12cm}} \def\Z{ {\rm Z \kern -.27cm \angle \kern
.02cm}} \def\N{ {\rm N \kern -.26cm \vrule width.4pt \kern .10cm}}
\def\1{{\rm 1\mskip-4.5mu l} }

\begin{document} \begin{center} 
\vbox to 1 truecm {}
{\large \bf The Isgur-Wise function in the BPS limit} \\

\vskip 1.5 truecm {\bf F. Jugeau}$^1$, {\bf A. Le Yaouanc}$^2$, {\bf L. Oliver}$^2\footnote{Speaker}$ {\bf and J.-C.
Raynal}$^2$\par \vskip 1 truecm

$^1$ Instituto de F\'isica Corpuscular (IFIC)\\
46071 Valencia, Spain
\par \vskip 5 truemm

$^2$ Laboratoire de Physique Th\'eorique\footnote{Unit\'e Mixte de
Recherche UMR 8627 - CNRS }\\    Universit\'e de Paris XI,
B\^atiment 210, 91405 Orsay Cedex, France 
\vskip 1 truecm
{\it  Talk given at the ICHEP 2006 Conference, Moscow, July 2006}
\end{center}
\vskip 1 truecm 

\begin{abstract} 
From sum rules in the heavy quark limit of QCD, using the non-forward amplitude, we demonstrate that if the slope $\rho^2 = - \xi '(1)$ of the Isgur-Wise function $\xi (w)$ attains its lower bound $\displaystyle{{3\over 4}}$ (as happens in the BPS limit proposed by Uraltsev), the IW function is completely determined, given by the function $\xi (w) = \left ({\displaystyle {2 \over w+1}}\right)^{3/2}$.
\end{abstract}

\vskip 2 truecm

\noi LPT Orsay 06-62 \par 
\noi October 2006
\par \vskip 1 truecm

\noindent e-mails : jugeau@ific.uv.es,
leyaouan@th.u-psud.fr, oliver@th.u-psud.fr 
\newpage \pagestyle{plain}
In the leading order of the heavy quark expansion of QCD, Bjorken sum rule (SR) \cite{1r,2r} relates the slope of the elastic Isgur-Wise (IW) function $\xi (w)$, to the IW functions of the transition between the ground state $j^P = {1 \over 2}^-$ and the $j^P = {1 \over 2}^+, {3 \over 2}^+$ excited states, $\tau_{1/2}^{(n)}(w)$, $\tau_{3/2}^{(n)}(w)$, at zero recoil $w=1$ ($n$ is a radial quantum number). This SR leads to the lower bound $-\xi '(1) = \rho^2 \geq {1 \over 4}$. A new SR was formulated by Uraltsev in the heavy quark limit\cite{3r}, involving also $\tau_{1/2}^{(n)}(1)$, $\tau_{3/2}^{(n)}(1)$, that implies, combined with Bjorken SR, the much stronger lower bound
\beq
\label{1e}
\rho^2 \geq {3 \over 4}
\eeq

A basic ingredient in deriving this bound was the consideration of the non-forward amplitude $B(v_i) \to D^{(n)}(v') \to B(v_f)$, allowing for general $v_i$, $v_f$, $v'$ and where $B$ is a ground state meson. In refs. \cite{4r,5r,6r} we have developed, in the heavy quark limit of QCD, a manifestly covariant formalism within the Operator Product Expansion (OPE), using the matrix representation \cite{7r} for the whole tower of heavy meson states \cite{8r}. We did recover Uraltsev SR plus a general class of SR that allow to bound also higher derivatives of the IW function, 
\beq
\label{3e}
(-1)^L\xi^{(L)}(1) \geq {(2L+1)!! \over 2^{2L}}
\eeq

The general SR obtained from the OPE can be written in the compact way \cite{4r}
\beq
\label{4e}
L_{Hadrons} (w_i, w_f, w_{if}) = R_{OPE} (w_i, w_f, w_{if})
\eeq

\noi where the l.h.s. is the sum over the intermediate $D$ states, while the r.h.s. is the OPE counterpart. This expression writes, in the heavy quark limit \cite{4r}~:
\bea
\label{5e}
&&\sum_{D=P,V}\sum_n Tr \left [ \overline{B}_f (v_f) \Gamma_f D^{(n)}(v')\right ]  Tr \left [ \overline{D}^{(n)} (v') \Gamma_i B_i(v_i)\right ] \xi^{(n)}(w_i) \xi^{(n)}(w_f)\nn \\
&&+ \hbox{ Other excited states}  = - 2 \xi (w_{if}) Tr \left [ \overline{B}_f (v_f) \Gamma_f P'_+ \Gamma_i B_i(v_i)\right ]
\eea

\noi where $w_i = v_i \cdot v' $, $w_f = v_f \cdot v'$,  $w_{if} = v_i \cdot v_f$. $P'_+ = \displaystyle{{1 + {/\hskip - 2 truemm v}' \over 2}}$ is the positive energy projector on the intermediate $c$ quark and the $B$ meson is the pseudoscalar ground state $(j^P, J^P) = \left ( {1 \over 2}^-, 0^-\right )$, where $j$ is the angular momentum of the light cloud and $J$ the spin of the bound state. The heavy quark currents considered in the preceding expression are $\overline{h}_{v'}\Gamma_i h_{v_i}$,  $\overline{h}_{v_f}\Gamma_f h_{v'}$ and $B(v)$, $D(v)$ are the $4 \times 4$ matrices representing the $B$, $D$ states \cite{7r,8r}. \par

The domain for the variables $(w_i, w_f, w_{if})$ is \cite{4r}~: $w_i \geq 1$, $w_f \geq 1$,
\beq
\label{8e}
w_iw_f - \sqrt{(w_i^2-1)(w_f^2-1)} \leq w_{if} \leq w_iw_f + \sqrt{(w_i^2 - 1) (w_f^2 -1)}
\eeq

In\cite{4r} the following SR were established. Taking $\Gamma_i = {/\hskip-2 truemm v}_i$ and $\Gamma_f = {/\hskip-2 truemm v}_f$ and $w_i = w_f = w$ one finds the so-called Vector SR
\bea
\label{10e}
&&(w+1)^2 \sum_{L\geq 0} {L+1 \over 2L+1}\ S_L (w,w_{if}) \sum_n \left [ \tau_{L+1/2}^{(L)(n)}(w)\right ]^2 \nn \\
&&+ \sum_{L\geq 1}  S_L (w,w_{if}) \sum_n \left [ \tau_{L-1/2}^{(L)(n)}(w)\right ]^2 = \left ( 1 + 2w + w_{if}\right ) \xi (w_{if})
\eea

\noi and for $\Gamma_i = {/\hskip-2 truemm v}_i\gamma_5$ and $\Gamma_f = {/\hskip-2 truemm v}_f\gamma_5$ one finds the Axial SR
\bea
\label{11e} 
&&\sum_{L\geq 0}  S_{L+1} (w,w_{if}) \sum_n \left [ \tau_{L+1/2}^{(L)(n)}(w)\right ]^2 +(w-1)^2 \sum_{L\geq 1}  {L \over 2L-1} S_{L-1} (w,w_{if})\nn \\
 && \sum_n \left [ \tau_{L-1/2}^{(L)(n)}(w)\right ]^2 = -  \left ( 1 - 2w + w_{if}\right ) \xi (w_{if})
\eea

\noi In the precedent equations the IW functions $\tau_{L\pm 1/2}^{(L)(n)}(w)$ correspond to the transitions ${1 \over 2}^- \to j = L \pm {1 \over 2}$ and the function $S_{L} (w,w_{if})$ is given by the Legendre polynomial
\beq
\label{12e}
S_L(w, w_{if}) = \sum_{0 \leq k \leq L/2} C_{L,k}\left ( w^2-1\right )^{2k} \left (w^2 - w_{if}\right )^{L-2k}
\eeq

\noi with 
\beq
\label{13e}
C_{L,k} = (-1)^k {(L!)^2 \over (2L)!}\ {(2L-2k)! \over k!(L-k)!(L-2k)!}
\eeq

Differentiating $n$ times both SR (\ref{10e}), (\ref{11e}) with respect to $w_{if}$ and going to the border of the domain $w_{if} = w = 1$, one gets, 
the bounds (\ref{3e}).\par

On the other hand, Uraltsev \cite{9r} has proposed a special limit of HQET, namely the so-called BPS limit, that implies $\rho^2 = {3 \over 4}$. We have demonstrated \cite{12NREF}, using the above SR, that if the slope reaches its lower bound (\ref{1e}), as happens in the BPS limit, then all derivatives reach their lower bounds (\ref{3e}), and then the Isgur-Wise function is completely fixed, namely
\beq
\label{17e}
\xi (w) = \left ( {2 \over w+1}\right )^{3/2}
\eeq

The motivation to introduce the BPS limit \cite{9r} has been the rather close values obtained from experiment in inclusive $B$ decay for the fundamental parameters $\mu_{\pi}^2$ and $\mu_G^2$~: 
\bea
\label{18e}
&&\mu_{\pi}^2 =\ - {<B(v)|O_{kin, v}^{(b)}|B(v)>\over 2m_B} \nn \\
&&\mu_G^2 =\ {<B(v)|O_{mag, v}^{(b)}|B(v)> \over 2m_B}
\eea

\noi i.e. the matrix elements of the operators that appear in the $1/m_Q$ perturbation of the HQET Lagrangian,
\bea
\label{19e}
&&O_{kin,v}^{(Q)} = \overline{h}_v^{(Q)}(iD)^2 h_v^{(Q)} \nn\\
&&O_{mag,v}^{(Q)} = {g_s \over 2} \ \overline{h}_v^{(Q)}  \sigma_{\alpha\beta} G^{\alpha\beta}h_v^{(Q)}
\label{20e}
\eea

\noi In terms of ${1 \over 2}^- \to {1 \over 2}^+$, ${3 \over 2}^+$ Isgur-Wise functions at zero recoil $\tau_j^{(n)}(1)$ and level spacings $\Delta E_j^{(n)}$ $(j = {1 \over 2}, {3 \over 2})$, these quantities read \cite{10r}
\bea
\label{21e}
&&\mu_{\pi}^2 = 6 \sum_n \left [ \Delta E_{3/2}^{(n)}\right ]^2 \left [ \tau_{3/2}^{(n)}(1)\right ]^2 + 3 \sum_n \left [ \Delta E_{1/2}^{(n)}\right ]^2 \left [ \tau_{1/2}^{(n)}(1)\right ]^2\\
&&\mu_{G}^2 = 6 \sum_n \left [ \Delta E_{3/2}^{(n)}\right ]^2 \left [ \tau_{3/2}^{(n)}(1)\right ]^2 - 6 \sum_n \left [ \Delta E_{1/2}^{(n)}\right ]^2 \left [ \tau_{1/2}^{(n)}(1)\right ]^2
\label{22e}
\eea

\noi The inequality $\mu_{\pi}^2 \geq \mu_G^2$ holds, and one has found empirically, from the inclusive decay  $\overline{B}_d \to X_c \ell \overline{\nu}_{\ell}$, that $\mu_{\pi}^2$ and $\mu_G^2$ are rather close \cite{11r} $\mu_{\pi}^2 \cong 0.4 \ {\rm GeV}^2$,   $\mu_G^2 \cong 0.35\ {\rm GeV}^2$.\par

The value of $\mu_G^2 \cong 0.35$~GeV$^2$ is obtained from the heavy-light mesons hyperfine splitting (see for example ref. \cite{9r}), while the value $\mu_{\pi}^2 \cong 0.4$~GeV$^2$ comes from the fit to inclusive $\overline{B}_d \to X_c \ell \overline{\nu}_\ell$ decay moments.\par 

Uraltsev has suggested a dynamical hypothesis that implements the limiting condition of $\mu_{\pi}^2$ and $\mu_G^2$ being equal, the so-called BPS approximation, $\mu_{\pi}^2 = \mu_G^2$.\par

Let us underline that our main purpose is a mathematical one within the heavy quark limit of QCD. Namely, the determination of the form of the Isgur-Wise function in the heavy quark limit by adding one dynamical assumption, the BPS condition. Let us consider the pseudoscalar $B$ meson at rest, $v = (1,0,0,0)$. The equation of motion of HQET in the heavy quark limit implies $iD^0h_v^{(b)}|B(v)>\ = 0$, where $D^{\mu}$ is the covariant derivative and $h_v$ is the heavy quark field. Uraltsev has proposed a new more specific constraint, valid only for the {\it pseudoscalar ground state} meson $B$, the so-called BPS constraint $\left ( \vec{\sigma}\cdot i \overrightarrow{D}\right ) h_v^{(b)}|B(v)>\ = 0$ that amounts to the vanishing of the smaller components of the heavy quark field {\it within the pseudoscalar} $B$ meson. It will be convenient in the following to write these two conditions in a covariant way, for any value of $v$. These equations then read,
\beq
\label{27e}
(iD\cdot v)h_v^{(b)}|B(v)>\ = 0\qquad  , \qquad \gamma_5 i\  {/\hskip - 3 truemm D} h_v^{(b)}|B(v) >\ = 0\nn 
\eeq

\noi From $i  \ {/\hskip - 3 truemm D}  i  \ {/\hskip - 3 truemm D}  = (iD)^2 + {g_s \over 2}  \sigma_{\alpha\beta} G^{\alpha\beta}$ this implies the equality $\mu_{\pi}^2 = \mu_G^2$.\par

Using the formalism of Leibovich et al. \cite{12r}, we have demonstrated in \cite{12NREF}, using translational invariance, the equations of motion and the BPS condition (\ref{27e}) that the slope and the curvature of the IW function satisfy~:
\beq
\label{NEQ}
- \xi '(1) = {3 \over 4}\ ,\ \xi '' (1) = {15 \over 16}
\eeq

\noi i.e. the lower bounds (\ref{3e}) are saturated.

One can demonstrate by induction that in general the $L$-th derivative attains its lower bound (\ref{3e}) 
\beq
\label{NEQ1}
(-1)^L \xi^{(L)}(1) = {(2L+1)!! \over 2^{2L}}
\eeq

\noi  We will assume the relation for $L-1$,
\beq
\label{55e}
(-1)^{L-1} \xi^{(L-1)}(1) = {(2L-1)!! \over 2^{2(L-1)}}
\eeq

\noi and use (\ref{10e}) and (\ref{11e}) to demonstrate for $L$.\par

Let us differentiate the SR (\ref{10e}), (\ref{11e}) $M$ times relatively to $w_{if}$. Using (\ref{12e})-(\ref{13e}), we need
\beq
\label{56e}
\left [ {\partial^M \over \partial w_{if}^M} S_L(w,w_{if})\right ]_{w_{if} = 1} = F_{L,M}(w)
\eeq

\noi where $F_{L,M}(w) =  R_{L,M}(w^2-1)^{L-M}$, with 
\bea
\label{58e}
 &&R_{L,M} = (-1)^M \sum_{0\leq k \leq (L-M)/2} (-1)^k\nn \\
 && {(L!)^2 \over (2L)!}\ {(2L-2k)! \over k!(L-k)!(L-M -2k)!}
\eea

\noi One obtains then two equations respectively for the vector and axial SR. In the Vector case we obtain two useful relations for $M=L$, $w = 1$ and for $M = L-1$ differentiating once relatively to $w$ and taking $w=1$. Similarly, for the Axial case we obtain three useful relations, taking $M=L-1$ and $w=1$, $M = L$ and $w=1$, and  $M = L$, differentiating once relatively to $w$ and making $w=1$.\par

To proceed with the proof by induction, we assume $\tau_{L-1-1/2}^{(L-1)(n)}(1)=0$, that implies (\ref{55e}). One obtains 
\bea
\label{72e}
 (-1)^{L} \xi^{(L)} (1)  & =&{(2L+1)!! \over 2^{2L}} + {2L+1 \over 4L}  L! \sum_n \left [ \tau_{L-1/2}^{(L)(n)}(1)\right ]^2\nn\\
&=& {(2L+1)!! \over 2^{2L}} + {4L^2-1 \over 4}  L! \sum_n \left [ \tau_{L-1/2}^{(L)(n)}(1)\right ]^2 
\eea

\noi that imply $\tau_{L-1/2}^{(L)(n)}(1) = 0$ and (\ref{NEQ1}), as we wanted to demonstrate. Since (\ref{NEQ1}) are the successive derivatives of (\ref{17e}), assuming natural regularity properties, in the BPS limit the Isgur-Wise function is given by expression (\ref{17e}).\par

In conclusion, we have demonstrated in this paper that if the heavy quark limit of QCD is supplemented with a dynamical assumption, namely the BPS approximation proposed by Uraltsev, the Isgur-Wise function is completely determined, given by the expression
\beq
\label{75e}
\xi (w) = \left ( {2 \over w+1}\right )^{3/2}
\eeq

This is a mathematical result that comes from the heavy quark limit of QCD plus the BPS condition introduced by Uraltsev. The comparison with data is not straightforward, since $1/m_Q$ and radiative corrections have not been taken into account. Indeed, the function that has to be extrapolated at $w=1$ to obtain $|V_{cb}|$ is the form factor $h_{A_1}(w)$, and moreover the two ratios of form factors $R_1(w)$, $R_2(w)$ are involved, that become $R_1(w) = R_2(w) = 1$ in the heavy quark limit, considered in this paper. In a recent BaBar paper, the fit to $h_{A_1}(w)$ gives a slope $\rho_{A_1}^2 = 1.14$ \cite{18r}. This is far away from the heavy quark limit result with the BPS condition $\rho^2 = 0.75$. However, to make a proper comparison, radiative corrections to the heavy quark plus BPS limit should be considered \cite{17NREF}, and the constraints on the slope from Voloshin SR, that result in an upper bound on $\rho^2$ that is close to the BPS limit, should also be taken into account \cite{18NREF}. This discussion deserves a delicate and detailed discussion that will be done elsewhere.  \par

In conclusion, we have obtained an explicit expression for the Isgur-Wise function $\xi (w)$ by implementing the heavy quark limit of QCD with a dynamical assumption, namely the BPS condition proposed by Uraltsev, coming from the condition $\mu_G^2 = \mu_{\pi}^2$ or, equivalently, from $\rho^2 = {3 \over 4}$.

\section*{Acknowledgments}
This work was supported by the EC contract HPRN-CT-2002-00311 (EURIDICE).

\end{document}